\def\year{2017}\relax
\documentclass[letterpaper]{article} 
\usepackage{aaai17}  
\usepackage{times}  
\usepackage{helvet}  
\usepackage{courier}  
\usepackage{url}  
\usepackage{graphicx}  
\begin{document}
%
\title{Value Alignment, Fair Play, and the Rights of Service Robots}
\author{Daniel Estrada\\
New Jersey Institute of Technology\\
djestrada@gmail.com\\
}
\maketitle
\begin{abstract}
Ethics and safety research in artificial intelligence is increasingly framed in terms of ``alignment'' with human values and interests. I argue that Turing's call for ``fair play for machines'' is an early and often overlooked contribution to the alignment literature. Turing's appeal to fair play suggests a need to correct human behavior to accommodate our machines, a surprising inversion of how value alignment is treated today. Reflections on ``fair play'' motivate a novel interpretation of Turing's notorious ``imitation game'' as a condition not of \textit{intelligence} but instead of \textit{value alignment}: a machine demonstrates a minimal degree of alignment (with the norms of conversation, for instance) when it can go undetected when interrogated by a human. I carefully distinguish this interpretation from the Moral Turing Test, which is not motivated by a principle of fair play, but instead depends on imitation of human moral behavior. Finally, I consider how the framework of fair play can be used to situate the debate over robot rights within the alignment literature. I argue that extending rights to service robots operating in public spaces is ``fair” in precisely the sense that it encourages an alignment of interests between humans and machines.
\end{abstract}

\section{Value Alignment and Turing's Test}
A substantial portion of contemporary research into ethics and artificial intelligence is devoted to the problem of ``value alignment'' (hereafter \textbf{VA}) \cite{allen2005artificial,yudkowsky2008artificial,yampolskiy2013artificial,soares2014aligning,russell2015research,arnold2017value}. Rather than deriving ethically appropriate action from first principles or from a direct recognition of the good, VA takes as its goal the (presumably simpler) task of designing AI that conforms to human values. AI that reliably conforms to human values is said to be ``aligned''. A primary concern in this literature is to establish methods that guarantee alignment, potentially within tight parameters, since it is argued that even small and seemingly innocuous cases of misalignment can quickly develop into a serious threat to general human safety \cite{yudkowsky2008artificial,bostrom2012superintelligent,babcock2016agi}.\par

There are reasons to be optimistic about VA as an approach to AI ethics, perhaps most significantly that the framework of ``alignment'' seems to lend itself to contemporary machine learning techniques like supervised learning \cite{mohri2012foundations}, where machines systematically improve their performance relative to a specified training set. There are also reasons to be skeptical that today's machine learning techniques are adequate for generating the complex forms of alignment required for participating in human moral communities \cite{arnold2017value}. However, rather than critiquing the VA literature directly, my goal in this paper is to reflect on connections between the discourse on value alignment and the historical discussion of Turing's notorious ``imitation game'', with the hopes that lessons from the latter might better inform our developing discussions of the former.\par

Turing's test, originally offered as an alternative to the question ``can machines think?'', has since become a standard benchmark for evaluating the intelligence of machines \cite{turing1950computing,saygin2000turing,copeland2004essential,copeland2017turing}. The test revolves around a comparison to human performance: if the machine cannot be correctly identified by a human interrogator after a few minutes of conversation, it is said to "pass" the test and can be called intelligent. The central criterion for passing the test is \textit{indistinguishability from human behavior} \cite{dretske1997naturalizing,saygin2000turing}. We might describe the demand for indistinguishability in terms of ``behavioral alignment'': a machine is behaviorially aligned just in case it behaves indistinguishably from a human. \cite{allen2000prolegomena} already recognized that since the set of moral behaviors is a proper subset of the total behaviors, what today is called ``value alignment'' can be interpreted as a special case of behavioral alignment. From this insight they propose a Moral Turing Test (\textbf{MTT}) \cite{allen2006machine,wallach2008moral,arnold2016against}. The MTT is passed by a machine that behaves indistinguishably from a human in a conversation about moral actions.\footnote{\cite{arnold2016against} argue that even a Total MTT variation, where evaluation of behaviors is not restricted to conversation but encompasses the full range of moral behaviors, is not sufficient for saving the MTT as a viable ethical criterion. For this reason, I will not dwell on the restriction to conversation behaviors in this paper. See \cite{harnad1989minds}.} Just as passing the original Turing Test is supposed to suggest a degree of intelligence on the basis of behavioral alignment, so too does passing the MTT suggest a degree of moral agency on the basis of moral alignment.\par 

Although the Turing Test is widely known and discussed, it is generally not accepted as a reliable test for intelligence. Criticisms of Turing's test abound in the literature, perhaps best summarized by Dretske: "despite indistinguishability, all is dark..." \cite{dretske1997naturalizing}. Critics worry that the mere imitation of human behavior is not sufficient for either intelligent or moral agency, and so Turing's test doesn't tell us what we want to know \cite{searle1980minds,dennett1981brainstorms,dreyfus1992computers}. Here the theoretical goals of the MTT come apart from those of VA. Researchers concerned about value alignment don't care whether the machine is a genuine moral agent; a pure automaton (like the Paperclip Maximizer \cite{bostrom2003ethical}) might still pose a threat to humanity if it is sufficiently misaligned. And conversely, a sufficiently aligned machine is no guarantee of moral agency, just as convincing automation is no guarantee of intelligent agency. For this reason, strong rejections of MTT sit awkwardly in the literature alongside expansive research programs into the constraints on alignment, even while the former is a clear example of the latter. For instance, \cite{arnold2016against} criticize the MTT as a standard for building moral machines, and yet they go on in \cite{arnold2017value} to develop some theoretical constraints on applying machine learning to value alignment, while drawing no strong connections between these discussions. The effect is to make it appear as if the MTT is either irrelevant or unhelpful to the to the discussion of value alignment. \par

Arnold et al. reject the MTT on several grounds, including that imitation cannot serve as the basis for intelligent moral agency. Echoing the traditional criticisms, they write, "What becomes ever clearer through explicating the conditions of an MTT is that its imitative premise sets up an unbridgeable gulf between its method and its goal" \cite{arnold2016against} The framework of an "unbridgeable gap" is familiar from the philosophy of mind \cite{dennett1991real}, and seems to render Turing's proposal inadequate for the task of developing genuine moral agents. However, if our task is not to develop intelligent moral agents \textit{per se} but merely to align our machines with our values, than the MTT may continue to prove useful. In the next section, I argue that Turing's principle of "fair play for machines" (\textbf{FP}) \cite{turing1947automatic} provides a non-imitative ground for evaluating the alignment of machines. I argue that the FP avoids many of the classic criticisms of Turing's test, and provides a satisfying method for applying Turing's insights to the problem of value alignment distinct from the MTT. 

\section{Fair Play for Machines}
\cite{proudfoot} points to a variety of sources in developing a rich, comprehensive account of the Turing test (``from every angle''). Special emphasis is given to his discussion of a ``little experiment'' \cite{turing1948intelligent} involving chess playing computers in an experimental design that is clearly the germ for his landmark 1950 paper. However, curiously missing from Proudfoot's analysis (and mentioned only in passing in \cite{leavitt2017}), is a short passage from the end of Turing's 1947 Lecture on the Automatic Computing Engine to the London Mathematical Society \cite{turing1947automatic,copeland2004essential,hodges2012alan}. Here Turing is also concerned with evaluating the performance and "I.Q." of chess-playing computers, which suggests this passage should be read alongside his 1948 and 1950 papers for a full appreciation of the developing proposal. Since it is so regularly overlooked, I quote Turing's argument below in full, with paragraph breaks and emphasis added: 

\begin{quotation}
``It might be argued that there is a fundamental contradiction in the idea of a machine with intelligence. It is certainly true that ‘acting like a machine’ has become synonymous with lack of adaptability. But the reason for this is obvious. Machines in the past have had very little storage, and there has been no question of the machine having any discretion. The argument might however be put into a more aggressive form. It has for instance been shown that with certain logical systems there can be no machine which will distinguish provable formulae of the system from unprovable, i.e. that there is no test that the machine can apply which will divide propositions with certainty into these two classes. Thus if a machine is made for this purpose it must in some cases fail to give an answer. On the other hand if a mathematician is confronted with such a problem he would search around a[nd] find new methods of proof, so that he ought eventually to be able to reach a decision about any given formula. This would be the argument.\par

\textbf{Against it I would say that fair play must be given to the machine.} Instead of it sometimes giving no answer we could arrange that it gives occasional wrong answers. But the human mathematician would likewise make blunders when trying out new techniques. It is easy for us to regard these blunders as not counting and give him another chance, but the machine would probably be allowed no mercy. In other words then, if a machine is expected to be infallible, it cannot also be intelligent. There are several mathematical theorems which say almost exactly that. But these theorems say nothing about how much intelligence may be displayed if a machine makes no pretense at infallibility. \par

To continue my plea for ‘fair play for the machines’ when testing their I.Q. A human mathematician has always undergone an extensive training. This training may be regarded as not unlike putting instruction tables into a machine. One must therefore not expect a machine to do a very great deal of building up of instruction tables on its own. No man adds very much to the body of knowledge, why should we expect more of a machine? Putting the same point differently, \textbf{the machine must be allowed to have contact with human beings in order that it may adapt itself to their standards}. The game of chess may perhaps be rather suitable for this purpose, as the moves of the machine’s opponent will automatically provide this contact.'' \cite{turing1947automatic,copeland2004essential}
\end{quotation}

There are many striking things to note about this passage. First, Turing is responding to a critic of the very idea of machine intelligence, whose argument points to some necessary (and therefore unbridgeable) gap between the performance of humans and machines. In this case, the critic appeals to G\"odel's incompleteness theorem \cite{godel1931formal,smullyan2001godel} as evidence of such a gap, an objection he returns to under the heading of ``The Mathematical Objection'' in \cite{turing1950computing}. Recall that Turing's major mathematical contribution \cite{turing1937computable} is the formal description of a ``universal computer'', which can in theory perform the work of any other computer. On my interpretation \cite{estrada2014rethinking}, the universality of his machines is what ultimately convinces Turing that computers can be made to think. Without any assumption of behaviorism or appeal to a principle of imitation, the syllogism runs as follows: if the brain is a machine that thinks, and a digital computer can perform the work of any other machine, then a digital computer can think. This syllogism is both valid and sound. However, Turing recognizes that G\"odel's  theorem shows ``that with certain logical systems there can be no machine which will distinguish provable formulae of the system from unprovable''. This straightforwardly implies that there are some things that even Turing's universal machines cannot do. This result does not invalidate the syllogism above. Still, Turing's critics draw an inference from (1) there are some things machines cannot do, to (2) humans can do things that (mere) machines cannot do. Although this inference is clearly invalid,\footnote{Turing's original response to the Mathematical Objection remains satisfying: ``The short answer to this argument is that although it is established that there are limitations to the powers of any particular machine, it has only been stated, without any sort of proof, that no such limitations apply to the human intellect.'' \cite{turing1950computing}} arguments of this form persist even among respected scholars today \cite{penrose1999emperor,floridi2016should}. The passage from his 1947 Lecture shows Turing contending with this perennial challenge several years before his formal presentation of the imitation game. In other words, Turing was clearly aware of  an ``unbridgeable gap'' objection, and both his ``little experiment'' and the principle of fair play serve as ingredients in his response. A full appreciation of Turing's position in this debate ought to take this evidence into account. \par

Second, the core of Turing's response is to offer a ``plea'' for what he calls ``fair play for machines''. This suggestion is proposed in the context of ``testing their I.Q.'', making explicit the connection between FP and the developing framework of Turing's test. Essentially, Turing is worried about a pernicious double standard: that we use one standard for evaluating human performance at some task, and a more rigorous, less forgiving standard for evaluating the machine's performances \textit{at the same task}. Other things equal, a double standard is patently unfair, and thus warrants a plea for "fair play". Of course, one might worry that the mere fact that the performance comes from a machine implies that other things \textit{aren't} equal. Since machines are different from humans, they ought to be held to different standards. But on my interpretation , Turing is primarily motivated by a conviction that universal computers can perform the work of any other machine, and so humans and computers are not essentially different. Turing's test isn't designed to prove that machines can behave like humans, since in principle this follows from the universality of the machines. Instead, the test is designed to strip the human evaluator of his prejudices against the machines, hence the call for fair play. \par

Notice that calling a standard of judgment unfair does not imply that the machines treated unfairly can ``think''. Therefore, FP alone cannot serve as a basis for evaluating the intelligence of machines in the style of the Turing Test. And indeed, Turing's argument makes clear that his appeal to FP is concerned not with the intelligence of the machine, but instead with the standards used to evaluate the machine's performance. After all, Turing's plea is made in defense of machines that are expected to be infallible, and whose performance might be compromised (by ``occasionally providing wrong answers'') in order to more closely approximate the performance of a human. Turing's point is that we'd never demand a human mathematician to occasionally make mistakes in order to demonstrate their intelligence, so it's strange to demand such performance from the machine. If Turing's test is motivated by a call for "fair play for machines", this should inform our interpretation of the test itself. Since the principle of fair play does not depend on an imitative premise, the rejection of Turing's test on this basis seems too hasty.\par

Finally, the quoted passage closes by highlighting the phrase ``fair play for machines'' again\footnote{It may be interesting to consider why the phrase "fair play for machines" doesn't appear in the 1950 paper. Many of the arguments from the passage appear in the final section of his 1950, under the subsection ``Learning Machines'', where he proposes building ``a mind like a child's'' in response to Lovelace's Objection \cite{estrada2014rethinking}. In this section he proposes a number of games to play with machines, including chess and twenty questions. He also expresses a worry that ``The idea of a learning machine may appear paradoxical to some readers.'' Turing's 1950's paper is self-consciously written to a popular audience; perhaps Turing worried that a plea for "fair play for machines", including those that aren't even intelligent, might also confuse his readers too much, and undermine the constructive argument he's given. Hopefully, readers 70 years later are not so easily confused.}, and arguing that ``the machine must be allowed to have contact with human beings in order that it may adapt itself to their standards.'' Clearly, Turing is approaching the challenge of evaluating the performance of machines, even in purely intellectual domains like chess, as a problem of behavioral alignment. Moreover, Turing argues that for machines to achieve that alignment, \textit{they must be allowed} certain privileges, in the interest of ``fair play''. Specifically, Turing argues that if we expect the machine to learn our standards, we must afford access to our behavior. In other words, he's arguing that constraints on \textit{human behavior} are necessary to achieve alignment: in how we evaluate and interact with out machines. This perspective is rare even in the alignment literature today, where concerns are overwhelmingly focused on how to constrain the machine to stay within bounds of acceptable human behavior.

More importantly, Turing suggests that we must be willing to interact with machines, even those that aren't intelligent, if we expect these machines to align to our standards. And  this is precisely the kind of interaction Turing's proposed imitation game encourages. These reflections open a new route to defending the importance of Turing's test in today's alignment literature. Turing's test is usually understood as a benchmark for intelligence, and the MTT as a benchmark for moral agency. Commentary traditionally recognizes Turing's worries about standards of evaluation \cite{saygin2000turing,arnold2016against}, but they interpret Turing's imitation game as attempting to settle on some specific standard of evaluation: namely, indistinguishability from human performance, or perfect imitation, as judged by another human. If the machine meets this standard, the machine is considered intelligent. We might call this a ``benchmark'' interpretation of Turing's test, or \textbf{BTT}. The MTT is an instance of BTT that sets the benchmark to imitate human moral behavior, for example. Many machine learning applications today will present themselves as meeting or exceeding human performance (at discrimination tasks, image recognition, translation, etc.), a legacy of Turing's influence on the field. Criticisms of Turing's test focus on whether this benchmark is appropriate for evaluating the machine's performance, with most concluding it is not an adequate measure of general intelligence. But the principle of fair play suggests Turing is less interested in setting a particular benchmark for intelligence, and more concerned with establishing that the standards for evaluation that are fair. Call this interpretation the Fair Play Turing Test \textbf{FPTT}. A machine passes the FPTT when it meets the same standards of evaluation used to judge human performance at the same task. On this interpretation, Turing's imitation game is meant to describe a scenario of "fair play" where the human biases against machines can be filtered out, and the machine could be judged in their capacity to carry a conversation by the same standards as any other human. We typically think of someone who can hold a conversation as being intelligent, so if a machine can also hold a conversation without being detected as non-human, we should judge it intelligent too. Not because conversation is some definitive marker of intelligence, as the BTT interpretation suggests, but rather because conversation is a standard that is often used to evaluate the intelligence of humans, and the principle of fair play demands holding machines to the same standards. On this interpretation, the sort of hostile interrogation typically seen in demonstrations of Turing's test \cite{aaronson2014} seems straightforwardly unfair, since we wouldn't expect an intelligent human to hold up well under hostile interrogation either. 

Since the principle of FP does not depend on imitation, the FPTT works in a subtly different way than the BTT. Passing the FPTT doesn't merely imply a machine performs at human levels; passing FPTT implies more strongly that the machine performs at these levels \textit{when evaluated by the same standards} used to judge human performance. For instance, we usually aren't skeptical of mere imitation when talking to a human, so raising this concern in the context of evaluating machine could signal a change in the standards of evaluation, and thus a violation of FP. Cases where machine performance is expected to diverge significantly from humans might warrant a multiplicity of standards. We might, for instance, expect driverless vehicles to adhere to more rigorous safety standards than we typically hold human drivers. Recognizing these misaligned standards as a violation of fair play doesn't necessarily imply the situation is unethical or requires correction. Instead, identifying a failure of fair play draws attention to the multiplicity of standards for evaluating a task, and the lack of a unifying, consistent framework for evaluating all agents at that task. The framework of FPTT easily extends to evaluating performance at tasks other than ``general intelligence'' where we are interested in consistent, unifying standards, including the task of moral alignment in particular contexts. \cite{arnold2016against} reject the MTT as a standard for evaluating moral agency on the basis of its imitative premise. But FPTT doesn't depend on an imitative premise, and only checks for alignment with the standards used to judge humans at a task. In the next section, I argue that this framework of fair play has direct application for evaluating the alignment of robots operating in our world. 

\section{The Rights of Service Robots}
Historically, the question of robot rights has turned on questions of personhood \cite{gunkel2012machine,bryson2017and}. Conditions on personhood typically involve both cognitive and moral attributes, such as "recognizing the difference between right and wrong" \cite{christman2008autonomy}. The consensus among scholars is that robots do not yet meet the conditions on minimal personhood, and will not in the near future. However, this consensus is inconclusive, and has been used to argue that robot rights might be necessary to protect machines that operate below the level of human performance \cite{darlingkate2012,darling2015s}. For instance, in 2017 San Francisco lawmakers implemented restrictions on “autonomous delivery services on sidewalks and public right-of-ways,” citing safety and pedestrian priority of use as motivating concerns \cite{joefitzgeraldrodriguez2017}. The proposal raises a natural question of whether these robots have the right to use public spaces, and to what extent a ban on robots might infringe on those rights. These questions seem independent of more general concerns about moral agency and personhood that typically frame the rights debate. Furthermore, it is well known that service robots operating in public spaces are typically subject to bullying and abusive behavior from the crowd (\cite{salvini2010design,salvini2010safe,brscic2015escaping}. Protecting robots from such treatment seems necessary independent of whether they meet strict conditions for personhood. \par

Like many cases of moral alignment, the case of the rights for service robots to operate on public sidewalks seems to demand a standard for evaluating the performance of machines that does not turn on any imitative comparison with human agents. Delivery robots do not have nor require the intellectual and moral capacities typical of humans; to compare their operation with human performance seems at best mismatched, at worst insulting. Interpreted as a benchmark of performance, these machines operate well below the threshold where Turing's test is relevant and the vocabulary of rights and personhood applies. In contrast to the benchmark interpretation, however, the principle of fair play suggests we look for standards of evaluation that are consistent across humans and machines. In the case of service robots, the focus of concern is on the nature of the task these robots are performing, and the standards already in use for evaluating such performances. There's an obvious comparison between service robots and service animals that is tempting, but I think is ultimately unhelpful. Importantly, animals feel pain and can suffer, and service animals are used to support persons with disabilities who can't otherwise access public resources. Service robots, in contrast, are used by tech companies to better service their clients, and it seems implausible that they can 'suffer' in a morally salient way. Given the distinct nature of these roles, holding service robots to the standards of service animals seems inappropriate. 

A closer analogy to the work of service robots can be found \cite{chopra2011legal}, who propose an alternative approach to robot law centered not on personhood but instead on a framework of legal agency. A legal agent is empowered to act on behalf of a principal, to whom the agent holds a fiduciary duty that contractually binds the agent to act on the principal's interest. For instance, a lawyer or accountant operates as a legal agent in the service of their clients. In the context of agency law, an agent's right to operate turns both on the capacities of the agent to faithfully represent the principal, and also on the nature and scope of the role being performed. The framework of agency law offers a systematic defense of robot rights which focuses legal and policy attention to the roles we want robots to play in our social spaces, and the constraints which govern operation of any agent performing these roles \cite{chopraestrada}. A social role analysis of robots \textit{as legal agents} has clear application to the protection of service robots operating in public spaces, including delivery robots and self-driving cars. But it also has natural extensions for the regulation of robots in a wide variety of other social roles, including robots that provide services in the context of law and justice, finances, transportation, education, socio-emotional support, sex work, public relations, and security. For instance, agency law provides a straightforward path to the regulation of bots on social media that are used to influence voters and elections \cite{ferrara2016rise}. From this perspective, social media bots are operating on behalf of their operators in the service of specific roles (campaign promotion, electioneering, etc), and therefore fall under the same legal frameworks that already exist to evaluate the ethics and legality of these activities. \par

The proposal to adopt robot rights grounded in a framework of legal agency deserves an explicit elaboration outside the scope of this paper. I raise the suggestion in this context to demonstrate how the principle of fair play might be used to guide developing standards for evaluating machine performances. Recall that the principle of fair play asks that we evaluate the machine according to the same standards used to judge the performance of a human at the same task. Thus, FPTT focuses our discussion on the task-specific standards for evaluation, rather than on the details of the performance of any particular machine, or the contrasts in performance across different agential kinds. In this way, FP also suggests an expansive research agenda for classifying and detailing the types of roles we might want robots to serve in, and the constraints on evaluating the performance of any agent filling that role. For instance, what should social media bots acting as campaign representatives be allowed to say or do? This is not an engineering question about the capabilities of any machine. It is a social policy question about what machines can and cannot do in the service of their role. If we want machines to align to our standards of performance, then Turing argues that fair play must be given to the machines. \par

Of course, we don't want every machine to align to our standards. If standards cannot be made consistent across humans and machines, it entails a stratification of the social order that divides humans from machines. One might worry that a social role defense of robot rights does not eliminate the stratification, but in fact imposes new social divides wherever there is a distinction in social roles, and so threatens a conception of rights grounded on the universal and inalienable rights of humanity. In this way, a social role defense might appear to be a kind of ``3/5th compromise for robots''. This worry is reinforced by a review of the history of agency law, which itself develops out of a logic of slavery and indentured servitude \cite{johnson2016status}. However, the social role analysis of agency law provides a way around this worry by laying out a direct path to full legal agency for robots. To say that robots serve as agents for principals does not preclude the robot from being a full legal agent, since obviously lawyers and accountants retain their personhood and agency even while acting as an agent for their principal. And of course, serving as a principal is yet another social role to perform, one with its own standards of evaluation. Making the the standards for principals explicit allows for a robot to serve as principal, first for other robots, perhaps as a manager or oversight for other robots acting on its behalf, and eventually to serve as a principal for itself, thus bridging the gap to full legal agency.

\section{Conclusions}
In this article we have reviewed some primary concerns of the value alignment literature, and shown these interests were present in the development of Turing's test as early as \cite{turing1947automatic}. We argued that a widespread rejection of Turing's test as a standard of intelligence has led scholars to overlook Turing's call for fair play as a source of inspiration in developing machines that are value-aligned. We have proposed an alternate interpretation of Turing's test which is inspired by Turing's call for ``fair play for machines'', and carefully distinguished this interpretation from benchmark interpretations like the Moral Turing Test. Finally we have briefly discussed how FPTT might be used to justify a defense of robot rights and sketch out a path to full agency on the basis of a social role analysis of agency law. 

\section{Acknowledgements}
Thanks to conversations with Samir Chopra, Jon Lawhead, Kyle Broom, Rebecca Spizzirri, Sophia Korb, David Guthrie, Eleizer Yudkowsky, Eric Schwitzgebel, Anna Gollub, Priti Ugghley, @eripsabot, richROT, the participants in my AI and Autonomy seminars and the Humanities department at NJIT, all my HTEC students at CTY:Princeton, and everyone in the \#botally and Robot Rights communities across social media, especially David Gunkel, Julie Carpenter, Roman V. Yampolskiy, Damien Patrick Williams, and Joanna Bryson. Thanks also to the organizers, participants, and tweeps at \#AIES.

\bibliography{value}
\bibliographystyle{aaai}
\end{document}